\documentclass[11pt]{article}

\usepackage{hyperref}
\usepackage{authblk}
\pdfoutput=1

\begin{document}
\title{Bubble Chandeliers}
\author[1]{Sigurdur T. Thoroddsen\thanks{sigurdur.thoroddsen@kaust.edu.sa}}
\author[1]{Marie-Jean Thoraval}
\author[2]{Kohsei Takehara}
\author[2]{Takeharu Goji Etoh}
\affil[1]{Division of Physical Sciences and Engineering \& Clean Combustion Research Center,
King Abdullah University of Science and Technology (KAUST),
Thuwal, 23955-6900, Saudi Arabia.}
\affil[2]{Department of Civil and Environmental Engineering,
Kinki University, Higashi-Osaka, Japan.}

\maketitle
\begin{abstract}
When a drop impacts at very low velocity onto a pool surface it is cushioned by a thin layer of air, 
which can be stretched into a hemispheric shape.
We use ultra-high-speed video imaging to show how this thin air-layer ruptures.
The number and locations of these ruptures determines the morphology of the resulting myriad of micro-bubbles.
This fluid dynamics video is submitted to the APS DFD Gallery of Fluid Motion 2012, part of the
65th Annual Meeting of the American Physical Society's Division of Fluid Dynamics (18-20 November, San Diego, CA, USA).
\end{abstract}

The video shows numerous video clips taken with an ultra-high-speed video camera capable of frame-rates up to 1 million fps.

\begin{enumerate}
\item In the experiments the impact is viewed from the side, within the original pool liquid.  
Most of the liquids were silicone oils of various viscosities, but we also show results using perfluorohexane.
The running numbers on the lower right corner in many of the videos clips, show the time in micro-seconds.
\item The work shown in the video parallels our recent publication:
\begin{itemize}
\item{Thoroddsen, S. T., Thoraval, M.-J., Takehara, K. and Etoh, T. G. (2012)
Micro-bubble morphologies following drop impacts onto a pool surface, 
{\it J. Fluid Mech.}, {\bf 708}, 469-479.}
\end{itemize} 
\end{enumerate}
\end{document}